\documentclass[a4paper,12pt]{article}
\usepackage{amssymb}
\begin{document}

\begin{titlepage}
\thispagestyle{empty}

\begin{flushright}
\end{flushright}
\vspace{5mm}

\begin{center}
{\Large \bf
Superalgebras from D-brane actions}
\end{center}
%\vspace{3mm}

\begin{center}
{\large D. T. Reimers}\\
\vspace{2mm}

\footnotesize{
{\it School of Physics, The University of Western Australia\\
Crawley, W.A. 6009, Australia}
} \\
{\tt  reimers@physics.uwa.edu.au}\\
\end{center}
\vspace{5mm}

\begin{abstract}
\baselineskip=14pt

The Noether charge algebras of D-brane actions contain two anomalous terms which modify the standard supertranslation algebra. We use a cocycle approach to derive associated spectra of topological charge algebras. The formalism is applied to $(p,q)$-strings and the D-membrane. The resulting spectra contain known algebras which allow the construction of extended superspace actions.

\end{abstract}

\vfill
\end{titlepage}

\renewcommand{\a}[0]{\alpha}
\renewcommand{\b}[0]{\beta}
\newcommand{\g}[0]{\gamma}
\newcommand{\G}[0]{\Gamma}
\renewcommand{\d}[0]{\delta}
\newcommand{\D}[0]{\Delta}
\newcommand{\e}[0]{\epsilon}
\renewcommand{\k}[0]{\kappa}
\newcommand{\m}[0]{\mu}
\newcommand{\n}[0]{\nu}
\newcommand{\p}[0]{\phi}
\renewcommand{\r}[0]{\rho}
\newcommand{\s}[0]{\sigma}
\renewcommand{\S}[0]{\Sigma}
\renewcommand{\t}[0]{\theta}
\newcommand{\x}[0]{\xi}

\newcommand{\lrb}[0]{\left (}
\newcommand{\rrb}[0]{\right )}
\newcommand{\lsb}[0]{\left [}
\newcommand{\rsb}[0]{\right ]}
\newcommand{\lcb}[0]{\left \{}
\newcommand{\rcb}[0]{\right \}}

\newcommand{\sa}[0]{\overrightarrow{\s}}
\newcommand{\nn}[0]{\nonumber}
\newcommand{\tb}[0]{\overline\theta}
\newcommand{\del}[0]{\partial}
\renewcommand{\bar}[0]{\overline}
\newcommand{\bea}[0]{\begin{eqnarray}}
\newcommand{\eea}[0]{\end{eqnarray}}

\newcommand{\half}[0]{\frac{1}{2}}
\newcommand{\third}[0]{\frac{1}{3}}
\newcommand{\sixth}[0]{\frac{1}{6}}
\newcommand{\dtgtl}[1]{d\overline{\theta}\Gamma_{#1}\theta}
\newcommand{\dtgtu}[1]{d\overline{\theta}\Gamma^{#1}\theta}

\newtheorem{theorem}{Theorem}

\section{Introduction}

Various types of branes are classified according to the CE (Chevalley-Eilenberg) cohomology \cite{chevalley48} of their field strengths. For $p$-branes, the WZ (Wess-Zumino) term is the pullback of a superspace form defined by its field strength. This field strength is the unique, nontrivial $(p+2)$-cocycle of the CE cohomology with the correct dimensionality \cite{azc89-2}. A similar classification also occurs for D-branes \cite{chrys99,sak00}. The CE nontriviality of these brane field strengths has some interesting consequences.

Firstly, the WZ term is necessarily super-Poincar\'{e} invariant only up to a total derivative. As a result, when the topology of the background superspace is nontrivial, the Noether charge algebra can be extended by a topological ``anomalous term" \cite{azc89}. For branes with worldvolume gauge fields, there is a second modification to the algebra that results from the transformation properties of the gauge field \cite{sorokin97,hammer97}. For D-branes, this modification is due to the presence of the BI (Born-Infeld) worldvolume gauge field. Terms of the D-brane Noether charge algebra associated with bosonic topology were explicitly found for the type IIA cases \cite{hammer97}. One must necessarily solve a series of descent equations to find the anomalous terms. Representative solutions to the D-brane descent equations were found, and the associated bosonic topological charges given \cite{hatsuda97,kamimura97,hatsuda98}.

There is a construction involving ghost fields which describes the appearance of anomalous terms in Noether charge algebras. In this construction, the anomalous term for the $p$-brane arises as an element of the second cohomology of a ``ghost differential" acting on a loop superspace \cite{azc91}. The appearance of anomalous terms in the D-brane Noether charge algebra can be described in a similar way \cite{Bergshoeff98}. The cohomological descent nature of the equations is manifest in this approach.

Extended superspace formulations have been considered in the case of both $p$-branes and D-branes. In the case of $p$-branes it was noted that extended superalgebras exist which allow manifestly super-Poincar\'{e} invariant WZ terms to be constructed \cite{siegel94,bergshoeff95,chrys99}. Extended superspace actions for $(p,q)$-strings, D-branes and string-brane systems can also be constructed \cite{sak98,sak99,chrys99,sak00,hatsuda01}. In all these cases, one seeks extended supertranslation algebras which trivialize the brane field strengths with respect to CE cohomology. Topological charge algebras of standard actions start to resemble these extended algebras once fermionic topological charges are considered. For example, superspaces which include both bosonic and fermionic topological charges can be candidates for the construction of extended superspace actions \cite{sezgin96}. In general, the bosonic topological charges now become noncentral. The explicit construction of fermionic charges was considered in \cite{hatsuda00,hatsuda01,peeters03}.

Recently, we approached $p$-brane topological charge algebras from the point of view of a single cocycle associated with the $p$-brane \cite{reimers05}. The WZ field strength and the anomalous term are described as two different representatives of this cocycle. Due to gauge transformations of the cocycle, the anomalous term is described as a full cohomology class. For the standard superspace action, this class is unique and nontrivial. Due to the gauge freedom, there is a full ``spectrum" of topological charge algebras resulting from the anomalous term. Upon retaining the terms associated with fermionic topology, the algebras used in extended superspace formulations of $p$-branes appear in the spectrum of topological charge algebras of the standard action \cite{reimers05,reimers05-2}.

In this paper, we generalize this work to the case of D-branes. There are two nontrivial cocycles associated with the D-brane, and each one generates an anomalous term of the Noether charge algebra. The topological charge algebras resulting from these anomalous terms are shown to be extensions of the standard supertranslation algebra by two disjoint, commuting ideals. Explicit representatives of both anomalous terms are found for the $(p,q)$-strings and the D-membrane. We generalize previous work in this regard by retaining the terms associated with fermionic topology. For the string, gauge freedom is used to generate a spectrum of topological charge algebras which is invariant under type IIB $SO(2)$ rotations. A topological charge algebra for $(p,q)$-strings is then deduced. For the membrane, the topological charge algebras associated with the NS-NS (Neveu-Schwarz) potential are derived. Although only the string and membrane algebras are explicitly derived, subalgebras associated with the NS-NS potential are common to all type IIB and type IIA D-branes respectively. In both cases, the spectrum of topological charge algebras contains known algebras which allow the construction of extended superspace actions.

The structure of this paper is as follows. In section \ref{sec:D-branes}, standard D-brane actions in flat backgrounds are reviewed. Two additional formulations of the action are then presented: a manifestly invariant formulation, and a set of $SO(2)$ dual actions for type IIB backgrounds. In section \ref{sec:D-brane cohomology}, the cocycle approach is generalized to D-branes. We review the relation between anomalous terms of the Noether charge algebra and the nontrivial cocycles of the D-brane. The single cocycle approach is presented. The resulting topological charge algebras are shown to be extensions of the standard supertranslation algebra by disjoint, commuting ideals. In section \ref{sec:Application to (p,q)-strings}, the general formalism is first applied to an $SO(2)$ dual set of D-strings. Representatives of the anomalous terms are found, and gauge freedom is then used to generate a spectrum of $SO(2)$ invariant topological charge algebras. A gauge fixed algebra for the $(p,q)$-strings is then deduced. In section \ref{sec:Application to the D-membrane}, representatives of the anomalous terms of the D-membrane are found. A spectrum of topological charge algebras associated with the NS-NS potential is then derived. In section \ref{sec:Comments}, we comment on the results.

\section{D-branes}
\label{sec:D-branes}

\subsection{Standard actions}

For this paper we will work with the standard, flat, background superspaces in d=10. The backgrounds are defined by the chirality of the spinors. Weyl spinors are eigenspinors of the idempotent ``chirality matrix:"
\bea
	\G_{11}=\G_0\ldots \G_9.
\eea
Since $\G_{11}$ is traceless, the eigenvalues are $\pm 1$ in equal numbers. Majorana spinors satisfy $\bar \t_\a=\t^\b C_{\b\a}$, where $C_{\b\a}$ is the antisymmetric charge conjugation matrix. Type IIA superspace consists of a single Majorana spinor (or equivalently, two Majorana-Weyl spinors of opposite chirality). Type IIB superspace consists of two Majorana-Weyl spinors of the same chirality. For type IIB superspace it will be assumed that spinor indices are accompanied by a suppressed index $I=(1,2)$ which identifies the spinor. The Pauli matrices $(\s_i)_{IJ}$ act upon these indices. Indices on Pauli matrices are raised and lowered with the Kronecker delta, while indices on gamma matrices are raised and lowered from the left by the charge conjugation matrix. $\G^a{}_{\a\b}$ is assumed to be symmetric. The de Rham differential acts from the right, and wedge product multiplication of forms is understood.

The superalgebra of the supertranslation group is:
\bea
    \{Q_{\alpha},Q_{\beta}\}&=&\Gamma^{a}{}_{\alpha\beta}P_{a}.
\eea
The corresponding group manifold can be parameterized:
\bea
    \label{2:parametrization}
    g(Z)&=&e^{x^{a}P_{a}}e^{\theta^{\a}Q_{\a}}\\
    Z^{A}&=&(x^{a},\theta^{\a}).\nn
\eea
The left vielbein is defined by:
\bea
	L(Z)&=&g^{-1}(Z)dg(Z)\\
	&=&dZ^{M}L_{M}{}^{A}(Z)T_{A}\nn,
\eea
where $T_A$ represents the full set of superalgebra generators. Its explicit components are:
\bea
    L^{a}&=&dx^{a}-\half \dtgtu{a}\\
    L^{\alpha}&=&d\theta^{\alpha}\nn.
\eea
The right vielbein is defined similarly:
\bea
	R(Z)&=&dg(Z)g^{-1}(Z)\\
	&=&dZ^{M}R_{M}{}^{A}(Z)T_{A}\nn.
\eea
The left action of the supertranslation group on itself is defined by:
\bea
	g(Z')&=&g(\e)g(Z).
\eea
This action is generated by operators $Q_A$ (``left generators"). One finds:
\bea
	\label{left generators for Z}
	\d Z^M&=&\e^AQ_{A}Z^M\\
	&=&\e^AR_{A}{}^{M}\nn,
\eea
where $R_{A}{}^{M}$ are the inverse right vielbein components, defined by:
\bea
	R_{A}{}^{M}R_{M}{}^{B}=\d_{A}{}^{B}.
\eea
Explicitly this yields:
\bea
\begin{array}{lll}
	&Q_\a x^m=-\half (\G^m\t)_\a,\qquad &Q_\a \t^\m=\d_\a{}^\m\\
	&Q_a x^m=\d_a{}^m,\qquad &Q_a \t^\m=0\nn.
\end{array}
\eea
Forms that are invariant under a global left action will be called ``left invariant." The left vielbein components are left invariant by construction.

Super-Dirichlet-$p$-branes (D$p$-branes) are $\k$-symmetric, $p+1$ dimensional manifolds (``worldvolumes") embedded in the background superspace. D$p$-branes in type IIA superspace exist only for $p$ even, while those in type IIB superspace exist only for $p$ odd. Actions for D-branes have been developed in both flat and more general backgrounds \cite{Schmidhuber96_D-brane_actions,aganagic97,cederwall96,Bergshoeff96}. We now present the action with the conventions adopted in this paper.

Let the worldvolume be parameterized by coordinates $\s^i$. The worldvolume metric $g_{ij}$ is defined using the pullbacks of the left vielbein components:
\bea
\label{worldvol metric}
    L_{i}{}^{A}&=&\del_{i}Z^{M}L_{M}{}^{A}\\
    g_{ij}&=&L_{i}}{^{a}L_{j}{}^{b}\eta_{ab}\nn.
\eea
The action consists of two terms:
\bea
	\label{standard action}
	S&=&S_{DBI}+S_{WZ}.
\eea
The DBI (Dirac-Born-Infeld) term is:
\bea
	\label{non manifest action}
	S_{DBI}=-\int d^{p+1}\s\sqrt{-\mathsf{det}(g_{ij}+F_{ij})}.
\eea
$F$ is a $2$-form\footnote{It suits us to have $dF=H$. Hence the difference in sign convention with respect to some prior literature.}:
\bea
	F&=&B-dA.
\eea
$A=d\s^iA_i$ is the BI worldvolume gauge field, which is a $1$-form defined only on the worldvolume. The NS-NS potential $B$ is a superspace $2$-form defined by:
\bea
	dB&=&H,
\eea
where $H$ is the left invariant, NS-NS $3$-form field strength. For type IIA superspace, $H$ is:
\bea
	H&=&\half L^a d\bar\t\G_{11}\G_ad\t,
\eea
while for type IIB:
\bea
	H&=&-\half L^a d\bar\t\G_a \s_3 d\t.
\eea
It is a characteristic feature of super-$p$-branes of various types that closure of field strengths requires ``Fierz identities" for products of gamma matrices. Closure of $H$ requires a ``standard" identity \cite{aganagic97}. For type IIA superspace this can be written:
\bea
	\G^a{}_{(\a\b}(\G_{11}\G_a)_{\g\d)}&=&0,
\eea
while for type IIB:
\bea
	\G^a{}_{(\a\b}(\G_a\s_3)_{\g\d)}&=&0.
\eea

The second term in the action is the WZ term:
\bea
	S_{WZ}&=&\int b.
\eea
It is defined by the formal sum of forms:
\bea
	b=\breve b e^{F}.
\eea
The form of degree $p+1$ is selected from this sum and the integral is then performed over the worldvolume of the brane. In general we will denote the form of a specific degree in a formal sum by a number in brackets. For example:
\bea
	\breve b&=&\oplus \breve b^{(n)}.
\eea
The R-R (Ramond) potentials $\breve b^{(n)}$ are defined by:
\bea
	\label{R-R potential defn}
	R=d\breve b +\breve b H.
\eea
The R-R field strengths $R^{(n)}$ are left invariant superspace forms:
\bea
	R^{(n)}=(-1)^pd\bar\t S^{(n-2)}d\t,
\eea
where for type IIA superspace the $S^{(n)}$ are given by:
\bea
	S^{(n)}&=&\frac{1}{2n!}L^{a_1}\ldots L^{a_n}\G_{a_1 \ldots a_n}\G_{11}{}^{[\frac{n}{2}+1]},
\eea
while for type IIB:
\bea
	S^{(n)}&=&\frac{1}{2n!}L^{a_1}\ldots L^{a_n}\G_{a_1 \ldots a_n}\s_3{}^{[\frac{n+1}{2}+1]}\s_1.
\eea
It follows from (\ref{R-R potential defn}) that the total field strength for the WZ term is the degree $p+2$ piece of:
\bea
	h&=&db\\
	&=&Re^F\nn.
\eea
Closure of $h$ is equivalent to some more general Fierz identities. For type IIA superspace these are:
\bea
	(m-1)(\G_{11}{}^{\frac{m}{2}}\G_{[a_1\ldots a_{m-2}})_{(\a\b}(\G_{11}\G_{a_{m-1}]})_{\g\d)}&&\\
		-\G^{a_m}{}_{(\a\b}(\G_{11}{}^{\frac{m+2}{2}}\G_{a_1\ldots a_m})_{\g\d)}&=&0,\nn
\eea
while for type IIB:
\bea
	(m-1)(\G_{[a_1\ldots a_{m-2}}\s_3{}^{\frac{m+1}{2}}\s_1)_{(\a\b}(\G_{a_{m-1}]}\s_3)_{\g\d)}&&\\
		+\G^{a_m}{}_{(\a\b}(\G_{a_1\ldots a_m}\s_3{}^{\frac{m+3}{2}}\s_1)_{\g\d)}&=&0.\nn
\eea
Most of these can be shown to hold by repeated use of the $m=2$ identity \cite{aganagic97,cederwall96}.

Left invariance of the action requires that the BI gauge field must transform under the left action of the supertranslation group. This transformation is determined by the requirement that the potential $F$ must be left invariant. Since $[d,Q_A]=0$, it is required:
\bea
	dQ_A A=Q_A B.
\eea
From the left invariance of $H$ it follows that:
\bea
	\label{Q_A B=dW_A}
	Q_A B&=&-dW_A
\eea
for some set of $1$-forms $W_A$. Hence:
\bea
	\label{Q_A A=-W_A}
	Q_A A_i=-(W_A)_i
\eea
is the required transformation of the BI gauge field \cite{hammer97}. Furthermore, since $H$ is CE nontrivial, there does not exist a potential $B$ such that $Q_A B=0$ for all $Q_A$ \cite{chrys99,sak00}.

\subsection{Manifestly left invariant action}

The variation of the WZ term of the standard action under the left group action is analogous to (\ref{Q_A B=dW_A}); from the left invariance of $h$ it follows that the variation of the WZ term is a total derivative:
\bea
	Q_A b=-dw_A.
\eea
Since $h$ is CE nontrivial, there does not exist a potential $b$ such that $Q_A b=0$ for all $Q_A$ \cite{chrys99,sak00}. As a result, the standard Lagrangian is not manifestly left invariant.

A manifestly left invariant formulation for D-branes which we will not explicitly describe here is the ``scale invariant" approach \cite{Bergshoeff98}. For the purposes of this paper we find it more convenient to define a simple, manifestly left invariant generalization of the standard action. First introduce an additional worldvolume $p$-form gauge field:
\bea
	a&=&d\s^{i_p}\ldots d\s^{i_1}a_{i_1\ldots i_p}\frac{1}{p!}
\eea
satisfying:
\bea
	\label{p form gauge field left trans}
	Q_A a_{i_1\ldots i_p}&=&-(w_A)_{i_1\ldots i_p}.
\eea
One then uses the alternative action:
\bea
	\label{manifest action}
	S&=&-\int d^{p+1}\s\sqrt{-\mathsf{det}(g_{ij}+F_{ij})}+\int f\\
	f&=&b-da\nn.
\eea
Unlike the components of the BI gauge field, the fields $a_{i_1\ldots i_p}$ are not physical degrees of freedom since they appear trivially (in a total derivative) in the action.

\subsection{Type IIB $SO(2)$ rotations}

There are various dualities relating different D-brane actions \cite{aganagic97-2}. If one includes nonvanishing background scalars (dilaton and axion) in the action, the dualities can be explicitly studied. Although this is an indirect issue for the purposes of this paper, in section \ref{sec:Application to (p,q)-strings} we will find it useful to consider the rotations of the type IIB D-string action. Classically there is an $SL(2,\mathbb{R})$ duality, but quantum considerations restrict this to $SL(2,\mathbb{Z})$. There is then an $SL(2,\mathbb{Z})$ multiplet of $(p,q)$-strings \cite{schwarz97,aganagic97-2,witten95,town97-S-duality,cederwall97}. Although the background scalars transform inhomogeneously under $SL(2,\mathbb{R})$, one may consistently set them to zero if one considers only the $SO(2)$ automorphism subgroup. The Pauli matrix $\s_2$ can be taken as the generator for these automorphisms, and the standard type IIB superspace action corresponds to a particular choice of $SO(2)$ frame \cite{aganagic97}. We wish to investigate how these frame rotations affect the results. The automorphisms can be implemented via rotations of the Pauli matrices \cite{cederwall96}. However, for studying the properties of the Noether charge algebra it is useful to have an implementation in terms of \textit{field transformations} instead. Such possibilities were considered in \cite{hatsuda97}. We take:
\bea
	x_\phi&=& x\\
	\t_\phi&=& e^{i\phi\s_2}\t\nn.
\eea
The worldvolume metric is invariant under these transformations. The worldvolume gauge field $A_{\phi}$ is defined as usual by its transformation properties (in particular, the left invariance of $F$ must be preserved). The set of type IIB D-brane actions $S_\phi$ with a free angular parameter $\phi$ is then:
\bea
	\label{rotated action}
	S_\phi[Z,A]=S[Z_\phi{},A_{\phi}].
\eea

\section{D-brane cohomology}
\label{sec:D-brane cohomology}

Using cohomological methods to investigate the anomalous terms of the Noether charge algebra gives insight into their geometrical origin. A constant ghost partner $e^A$ is introduced for each superspace coordinate. A ``generalized" $(m,n)$-form $Y$ is then written:
\bea
	Y&=&e^{B_{n}}\ldots e^{B_{1}}L^{A_{m}}\ldots L^{A_{1}}Y_{A_{1}\ldots A_{m},B_{1}\ldots B_{n}}\frac{1}{m!n!}.
\eea
The space of $(m,n)$-forms will be denoted $\Omega^{m,n}$, and the collection of such spaces $\Omega^{*,*}$. Because D-branes have worldvolume forms that cannot be defined on the background superspace, the space $\Omega^{m,n}$ will consist of worldvolume forms. Where a superspace form is used in the construction, the pullback of that form to the worldvolume is implied. A ghost differential $s$ introduced in \cite{azc91} can be defined by the properties:
\begin{itemize}
\item
$s$ is a right derivation. That is, if $X$ and $Y$ are generalized forms and $n$ is the ghost degree of $Y$ then:
\bea
    s(XY)=Xs(Y)+(-1)^{n}s(X)Y.
\eea
\item
If $X$ has ghost degree zero then:
\bea
    sX=e^{A}Q_AX.
\eea
\item
\bea
    se^{A}=\half e^{C}e^{B}t_{BC}{}^{A},
\eea
where $t_{BC}{}^A$ are the structure constants of the supertranslation algebra.
\end{itemize}
The operators $s$ and $d$ commute. However, for $\{s,d,\Omega^{*,*}\}$ to define a double complex we must show that $s$ is nilpotent (i.e. $s^2=0$). In the case of $p$-branes where everything is defined on the background superspace, this turns out to be identically true. For D-branes, the transformation properties of the BI gauge field (which is not part of the background) complicate the issue. Nilpotency of $s$ does not hold for an action on arbitrary fields (for example $s^2A\neq 0$). However, the BI gauge field appears in the action only through the potential $F$. One of the defining properties of $F$ is its left invariance, which may be written \cite{Bergshoeff98}:
\bea
	sF&=&0.
\eea
It follows that $s$ is nilpotent (and defines a double complex) when we restrict the BI gauge field to appear in $\Omega^{*,*}$ only through $F$.

The total differential $D$ is \cite{reimers05}:
\bea
    D&=&s+(-1)^{n+1}d\\
    D^{2}&=&0\nn.
\eea
Generalized $l$-forms are defined on an associated single complex $\Omega^*_{D}$, which is the anti-diagonal of the double complex:
\bea
    \Omega_{D}^{l}=\{\oplus\Omega^{m,n}:\quad m+n=l\}.
\eea
The $l$-th cohomology of $D$ is:
\bea
    H_D^{l}=Z_D^{l}/B_D^{l},
\eea
where $Z_D^{l}$ are the $D$ cocycles, and $B_D^{l}$ are the $D$ coboundaries. The restriction of $H_D^l$ to representatives within $\Omega^{m,l-m}$ will be denoted $H^{m,l-m}$. The $D$ cocycle of the $p$-brane is associated with the CE nontrivial (p+2)-form field strength of the WZ term. The D-brane has two such field strengths: the NS-NS $3$-form $H$ and the WZ $(p+2)$-form $h$. As a result, there are two separate $D$ cocycles associated with the D-brane: the ``NS-NS cocycle" and the ``WZ cocycle".

First consider the NS-NS field strength $H=dB$. This is a nontrivial element of the CE cohomology in both the IIA and IIB cases \cite{chrys99,sak00}. The $D$ cocycle associated with $H$ exists in the ``NS-NS double complex." All elements of this complex are required to be Lorentz invariant, generalized forms of dimension two. The commuting nature of the operators leads to the descent equations \cite{Bergshoeff98}:
\begin{figure}[t]
\begin{center}
\begin{picture}(120,130)(0,-10)
\put(0,18){\vector(1,0){123}}
\put(0,18){\vector(0,1){85}}
\put(10,10){\makebox(50,50){
\large $\begin{array}{cccccccc}
& 3\ & dB & & &\\
& 2\ & B &\diamondsuit& &\\
\uparrow & 1\ & & W &\diamondsuit&\\
d & 0\ & & & N & sN\\
& & 0 & 1 & 2 & 3\\
& & s & \rightarrow & &
\end{array}$
}}
\end{picture}
\caption{Descending sequence for the NS-NS field strength}
\label{fig:NS-NS cocycle}
\end{center}
\end{figure}
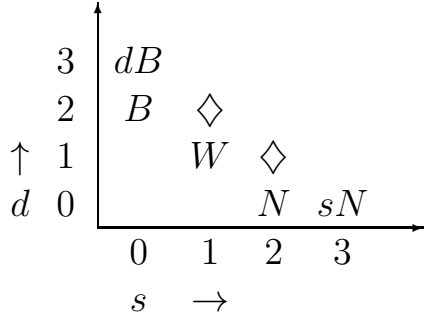
\bea
	H&=&dB\\
	sB&=&-dW\nn\\
	sW&=&dN\nn,
\eea
which are graphically depicted in the ``tic-tac-toe box" \cite{bott82} of figure \ref{fig:NS-NS cocycle}. The different representatives of the NS-NS cocycle are found on the LHS of these equations. Just as in the $p$-brane case, there is gauge freedom for the cocycle \cite{reimers05}. The gauge fields for the NS-NS cocycle that are of interest to us are $\Psi\in\Omega^{1,0}$ and $\Lambda\in\Omega^{0,1}$. The corresponding transformations can be summarized as:
\bea
	\D(B\oplus W\oplus N)=D(\Psi\oplus \Lambda).
\eea
Explicitly this gives:
\bea
	\label{gauge transform of potentials}
	\D B&=&-d\Psi\\
	\D W&=&s\Psi+d\Lambda\nn\\
	\D N&=&s\Lambda\nn.
\eea

Gauge transformations of the BI gauge field follow from those of the cocycle potentials. We now derive these transformations. The left transformation (\ref{Q_A A=-W_A}) of $A$ can be written \cite{Bergshoeff98}:
\bea
	\label{sA=-W}
	sA&=&-W.
\eea
However, due to the gauge transformations (\ref{gauge transform of potentials}), the potential $W$ is not unique. Equation (\ref{sA=-W}) then implies that the BI gauge field must transform under $\Psi$ and $\Lambda$. Firstly, the left invariant potential $F$ should be gauge invariant in order to preserve the symmetries of the action. By requiring invariance of $F$ under $\Psi$ it follows that the general form for the gauge transformations of $A$ is:
\bea
	\D A&=&-\Psi-d\Upsilon.
\eea
However, invariance of $F$ under $\Lambda$ means that the gauge fields $\Upsilon$ and $\Lambda$ are not independent; they must be related by:
\bea
	s\Upsilon&=&\Lambda.
\eea
In general this has no solution if $\Upsilon$ is a scalar on the background superspace. Therefore $\Upsilon$ must be a worldvolume scalar. Note that this is analogous to the interpretation of the BI gauge field. The equation (\ref{sA=-W}) has no solution if $A$ is defined as a superspace form (the nontriviality of $H$ prevents such a solution); therefore $A$ must be a new degree of freedom defined on the worldvolume.

The algebra of conserved charges of the D-brane action contains an anomalous term due to the transformation properties of the BI gauge field \cite{sorokin97,hammer97}. Let $(P_M,P^i)$ denote the momenta conjugate to $(Z^M,A_i)$. The minimal charges of the action are:
\bea
	\label{minimal Noether charges}
	\bar Q_A&=&\int d^p\s\Big [Q_A Z^M P_M+Q_A A_i P^i\Big ]\\
	&=&\int d^p\s\Big [R_A{}^M P_M-(W_A)_i P^i\Big ]\nn,
\eea
where the integral is over the spatial section of the worldvolume. Introduce the fundamental (graded) Poisson brackets for the phase space\footnote{Different types of bracket operation are used in this paper. We will not explicitly indicate the type since this should be clear within context.}:
\bea
	\lsb P_{M}(\s),Z^{N}(\s')\rcb &=&\d_{M}{}^{N}\d(\sa-\sa')\\
	\lsb P^i(\s),A_j(\s')\rcb &=&\d^i{}_j\d(\sa-\sa'),\nn
\eea
where it is assumed $\s'^{0}=\s^{0}$ (i.e. equal time brackets). The Dirac delta function notation is shorthand for the product of the $p$ delta functions associated with the spatial coordinates of the worldvolume. Let us denote the $H^{1,2}$ cocycle representative by:
\bea
M&=&sW.
\eea
One then obtains the ``minimal algebra" under Poisson bracket \cite{hammer97}:
\bea
	\label{minimal algebra no hat}
	\lsb \bar Q_A,\bar Q_B\rcb&=&-t_{AB}{}^C \bar Q_C-\int d^p\s (M_{AB})_iP^i.
\eea
For convenience we define a ``hat map" for elements $Y\in\Omega^{1,n}$ of the NS-NS double complex:
\bea
	\hat Y&=&-\int d^p\s Y_iP^i,
\eea
so that the algebra (\ref{minimal algebra no hat}) is:
\bea
	\lsb \bar Q_A,\bar Q_B\rcb &=&-t_{AB}{}^C \bar Q_C+\hat M_{AB}.
\eea
The minimal algebra is therefore already a modification by $\hat M_{AB}$ of the standard supertranslation algebra due to the presence of the BI gauge field. This modification will be referred to as the ``NS-NS anomalous term" (since it descends from the NS-NS field strength $H$).

The BI gauge field appears in the action only through its field strength. This leads to constraints on the conjugate momenta $P^i$ \cite{hammer97}. Firstly, since $\frac{\del\mathcal{L}}{\del(\del_0A_0)}=0$, there is the primary constraint:
\bea
	P^0&=&0.
\eea
Denote the spatial worldvolume coordinates by $\s^I$. The Euler-Lagrange equation for $A_0$ then yields the secondary ``Gauss law" constraint:
\bea
	\del_{I}P^{I}&=&0.
\eea
Now applying these constraints, and using $M=dN$, gives:
\bea
	\hat M_{AB}&=&-\int d^p\s \del_I (N_{AB}P^I).
\eea
The NS-NS anomalous term therefore consists of topological integrals, just as the $p$-brane anomalous term does. Note that once the constraints are imposed, the minimal charges lose their status as generators of the left group action. Therefore, the constraints should be applied only after the topological charge algebra has been evaluated.

Just as in the case of the $p$-brane, the minimal charges (\ref{minimal Noether charges}) are generally non-conserved, and this is due to quasi-invariance of the WZ term {\cite{azc89,hammer97}}. The second modification to the Noether charge algebra derives from the WZ field strength. The first three descent equations for the fields of the ``WZ double complex" are:
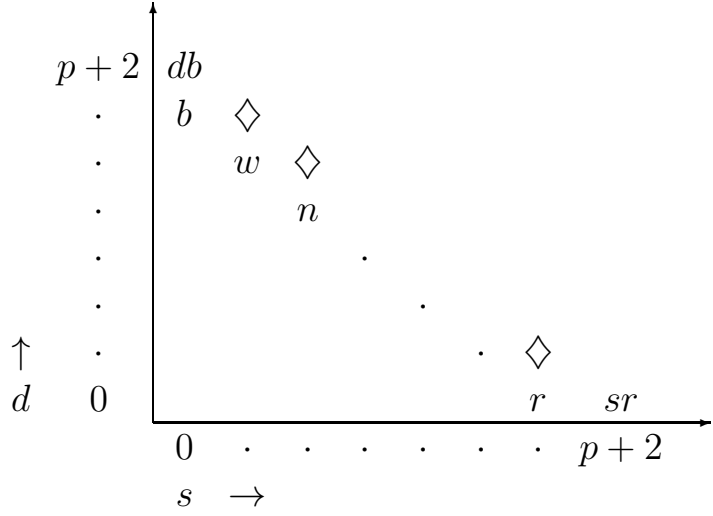
\begin{figure}[t]
\begin{center}
\begin{picture}(140,190)(0,-20)
\put(-35,12){\vector(1,0){211}}
\put(-35,12){\vector(0,1){159}}
\put(10,40){\makebox(50,50){
\large $\begin{array}{ccccccccccc}
& p+2 & db & & &\\
& \cdot & b &\diamondsuit& &\\
& \cdot & & w &\diamondsuit&\\
& \cdot & & & n &&&&& \\
& \cdot &&&&\;\cdot\;&&&& \\
& \cdot &&&&&\;\cdot\;&&& \\
\uparrow & \cdot &&&&&&\;\cdot\;&\diamondsuit&\\
d & 0 &&&&&&& r & sr\\
& & 0 & \cdot & \cdot & \cdot & \cdot & \cdot & \cdot & p+2 \\
& & s & \rightarrow & &
\end{array}$
}}
\end{picture}
\caption{Descending sequence for the WZ field strength}
\label{fig:WZ cocycle}
\end{center}
\end{figure}
\bea
	h&=&db\\
	sb&=&-dw\nn\\
	sw&=&dn\nn.
\eea
The sequence ends with the potential $r\in\Omega^{0,p+1}$, and the associated cocycle representative $sr\in H^{0,p+2}$. This has been depicted in the tic-tac-toe box of figure \ref{fig:WZ cocycle}. The exponential $e^F$ in the WZ term is preserved by the operators $d$ and $s$. All fields of the sequence are therefore formal sums containing this factor. Defining:
\bea
	w=\breve w e^F,
\eea
the descent equation $sb=-dw$ is then equivalent to \cite{Bergshoeff98}:
\bea
	\label{sb descent equation}
	s\breve b=-\breve w H-d\breve w.
\eea
The $H^{p,2}$ cocycle representative is then:
\bea
	\label{WZ anomalous term m}
	m&=&\breve m e^F\\
	&=&s\breve w e^F\nn.
\eea
This leads to the algebra of conserved charges as follows. The variation of the WZ term is a total derivative:
\bea
	Q_A \mathcal{L}_{WZ}&=&-\del_iw_A{}^i,
\eea
where
\bea
	\label{w_A^i def}
	w_{A}{}^{i}&=&\frac{1}{p!}\tilde \e^{i_{p}\ldots i_{1}i}w_{i_{1}\ldots i_{p},A}.
\eea
The conserved currents associated with this quasi-invariance are then:
\bea
	\label{conserved currents}
	\overrightarrow Q_{A}{}^{i}&=&Q_{A}Z^{M}\frac{\del \mathcal{L}}{\del(\del_{i}Z^{M})}
		+Q_{A}A_j\frac{\del \mathcal{L}}{\del(\del_{i}A_j)}
		+w_{A}{}^{i}\\
	\del_{i}\overrightarrow Q_{A}{}^{i}&=&0\nn.
\eea
Let the spatial section of the worldvolume be a closed manifold embedded in superspace by the map $\Phi$. For convenience we define a ``bar map" by its action on $(p,n)$-forms $Y$:
\bea
	\label{bar map}
	\bar Y&=&(-1)^p\int \Phi^* Y.
\eea
The conserved charges of the currents (\ref{conserved currents}) are then ``modified Noether charges:"
\bea
	\widetilde {\bar Q}_A&=&\bar Q_A+\bar w_A.
\eea
The $\widetilde {\bar Q}_{A}$ obey a modified version of the minimal algebra \cite{hammer97}:
\bea
	\label{modified Noether algebra}
	\lsb \widetilde {\bar Q}_{A},\widetilde {\bar Q}_{B}\rcb&=&-t_{AB}{}^{C}\widetilde {\bar Q}_{C}+\hat M_{AB}+\bar m_{AB},
\eea
with
\bea
	\bar m_{AB}=\lsb\bar Q_{A},\bar w_{B}\rcb+\lsb\bar w_{A},\bar Q_{B}\rcb+t_{AB}{}^{C}\bar w_{C}.
\eea
We refer to $\bar m$ as the ``WZ anomalous term" (since it descends from the WZ field strength $h$). Just as in the $p$-brane case, the components $\bar m_{AB}$ are topological integrals since $m=dn$ is a closed form.

Let us investigate what happens if we use the manifestly left invariant action (\ref{manifest action}) instead of the standard one. In this case there will be no contribution to the topological charge algebra from quasi-invariance of the WZ term. However, the mechanism outlined for the BI gauge field contribution also applies to the $p$-form gauge field \cite{hammer97,Bergshoeff98}. Since the worldvolume is $p+1$ dimensional, before constraints are taken into account, the $p$-form gauge field has $p+1$ independent components. We will conveniently take\footnote{Note this is the same ``Hodge dual like" map used in (\ref{w_A^i def}).}:
\bea
	\label{a^i def}
	a^i&=&\frac{1}{p!}\widetilde \e^{i_p\ldots i_1i}a_{i_1\ldots i_p}
\eea
as the independent components. The left transformation of $a^i$ follows from (\ref{p form gauge field left trans}) and (\ref{w_A^i def}):
\bea
	Q_Aa^i&=&-w_A{}^i.
\eea
Define the momenta conjugate to $a^i$:
\bea
	p_i&=&\frac{\del}{\del (\del_0a^i)}\mathcal{L}.
\eea
The conserved charges are then the Noether charges:
\bea
	\bar Q_A&=&\int d^p\s[Q_A Z^M P_M+Q_A A_i P^i+Q_A a^i p_i]\\
	&=&\int d^p\s[R_A{}^M P_M-(W_A)_i P^i-w_A{}^ip_i]\nn.
\eea
This yields the Noether charge algebra:
\bea
	\label{manifest charge algebra}
	\Big [\bar Q_A,\bar Q_B\Big \}&=&-t_{AB}{}^C \bar Q_C-\int d^p\s \Big [(M_{AB})_iP^i
		+m_{AB}{}^ip_i\Big ],
\eea
with $m_{AB}{}^i$ defined in the same way as (\ref{w_A^i def}), (\ref{a^i def}). These charges are once again topological in nature as a result of constraints for the momenta conjugate to the $p$-form gauge field. These constraints, which arise in the same way as those for the BI gauge field, are found to be:
\bea
	\del_{I}p_0&=&0\\
	p_I&=&0\nn.
\eea
In fact, since the $p$-form gauge field enters the action (\ref{manifest action}) trivially, we can simply evaluate the momenta to obtain:
\bea
	\label{p-form momenta value}
	p_0&=&-1\\
	p_I&=&0\nn.
\eea
Using this in (\ref{manifest charge algebra}), we then recover exactly the topological charge algebra (\ref{modified Noether algebra}) of the standard action, but with $\widetilde {\bar Q}_A$ replaced by $\bar Q_A$. That is, the conserved charges are now strict Noether charges instead of ``modified" ones. Thus, whether one uses the standard action (\ref{standard action}) or the manifestly invariant one (\ref{manifest action}), the algebra of conserved charges is the same. This is essentially the result suggested in \cite{Bergshoeff98} for the scale invariant formulation, although with a minor difference. In the scale invariant formulation, the $p$-form momenta $p_0$ is not fixed to a specific value as in (\ref{p-form momenta value}), so it becomes a constant multiplying the associated anomalous term. The same observations also clearly apply to the analogous formulations of ordinary $p$-brane actions.

For $p$-branes, the topological charge algebras can be analyzed in terms of operators and forms based in the double complex \cite{reimers05}. The anomalous term is thus seen to generate an extension of the background superalgebra by an ideal. We now show that this procedure also applies to the anomalous terms of the D-brane Noether charge algebra. For the WZ anomalous term (\ref{WZ anomalous term m}), the only difference from the case of the $p$-brane is the presence of factors of $F$. However, since $F$ is left invariant, only the variations of $\breve m$ contribute to the algebra. For the NS-NS anomalous term, the additional feature is the presence of the momenta $P^i$ conjugate to the components of the BI gauge field. At first this seems to complicate matters since both the conserved charges $\widetilde {\bar Q}_A$ and the WZ anomalous term $\bar m$ have dependence upon $A_i$. This could in principle generate ``cross terms" that do not arise in the case of the $p$-brane (because there are no momenta in the $p$-brane anomalous term). However, it turns out that these cross terms vanish. Firstly, $A_i$ appears only through its field strength, in products of:
\bea
	F_{ij}&=&B_{ij}-2\del_{[i}A_{j]}.
\eea
Using the bracket:
\bea
	[F_{ij}(\s),P^k(\s')]&=&-2\d_{[i}{}^k\del_{j]}\d(\s-\s'),
\eea
one then finds:
\bea
	[F_{ij}(\s),\hat M_{AB}]&=&-2\del_{[i}\del_{j]}N_{AB}(\s)\\
	&=&0\nn.
\eea
If $M_{AB}$ is split into closed forms representing superalgebra generators, the same calculation also holds for each generator. We thus have:
\bea
	\ [\bar m_{AB},\hat M_{CD}\}&=&0.
\eea
It also follows that the action of the conserved charges on the anomalous terms is equivalent to the action of the minimal charges:
\bea
	\Big [\widetilde {\bar Q}_A,\hat M_{CD}\Big \}&=&\Big [\bar Q_A,\hat M_{CD}\Big \}\\
	\Big [\widetilde {\bar Q}_A,\bar m_{CD}\Big \}&=&\Big [\bar Q_A,\bar m_{CD}\Big \}\nn.
\eea
The result is that we may use the double complex to find the topological charge algebra. Define the ``modified left generators:"
\bea
	\widetilde Q_A&=&Q_A+w_A.
\eea
We assign to $Q_A$ the minimal algebra:
\bea
	\lsb Q_A,Q_B\rcb&=&-t_{AB}{}^C Q_C+M_{AB}.
\eea
One then finds that the algebra generated by $\widetilde Q_A$ and $\{M_{AB},m_{AB}\}$ (taking forms to commute with forms) is the same as that generated by $\widetilde {\bar Q}_A$ and $\{\hat M_{AB},\bar m_{AB}\}$ under Poisson brackets. We may thus find the topological charge algebra by using the forms which represent the two anomalous terms. It also allows the spectrum of algebras resulting from each anomalous term to be considered in isolation. We summarize with:
\begin{theorem}[extension]
The anomalous terms of the D-brane Noether charge algebra define extensions of the standard supertranslation algebra by two disjoint ideals. The first derives from the cohomology class $[M]\in H^{1,2}$ of representatives for the NS-NS cocycle, the second from the class $[m]\in H^{p,2}$ of representatives for the WZ cocycle. The generators of both ideals commute amongst themselves and with each other.
\end{theorem}

\section{Application to $(p,q)$-strings}
\label{sec:Application to (p,q)-strings}

\subsection{D-strings}
\label{sec:D-strings}

Let us investigate a combination of the manifestly left invariant action (\ref{manifest action}) and the rotated action (\ref{rotated action}). The SO(2) rotated fields are given by\footnote{Note that $e_\phi{}^{\a I}$ are chiral ghost fields while $(e^{i\phi\s_2})^I{}_J$ is an exponential.}:
\bea
	\label{SO(2) field transformations}
	x_\phi{}^a &=&x^a\\
	\t_\phi{}^{\a I} &=& (e^{i\phi\s_2})^I{}_J\t^{\a J}\nn\\
	e_\phi{}^a &=&e^a\nn\\
	e_\phi{}^{\a I} &=& (e^{i\phi\s_2})^I{}_Je^{\a J}\nn\\
\left [\begin{array}{c}
A_\phi \\
a_\phi
\end{array}\right ]
&=&
\left [\begin{array}{cc}
\cos(2\phi) & -\sin(2\phi) \\
\sin(2\phi) & \cos(2\phi)
\end{array} \right ]
\left [\begin{array}{c}
A \\
a
\end{array}\right ]\nn.
\eea
The string case is special in that the field strengths $(H,h)$ transform as an $SO(2)$ vector doublet.
The potentials and worldvolume gauge fields of the double complex will be chosen such that they respect this transformation property. That is, only solutions to the descent equations which transform as vector doublets under (\ref{SO(2) field transformations}) will be considered. The defining properties of the relevant doublets are:
\begin{itemize}
\item
$(B,b)$:
\bea
	dB&=&H\\
	&=&\half L^ad\bar\t\G_a\s_1d\t\nn\\
	db&=&h\nn\\
	&=&\half L^ad\bar\t\G_a\s_3d\t\nn.
\eea
\item
$(W,w)$:
\bea
sB&=&-dW\\
sb&=&-dw\nn.
\eea
\item
$(A,a)$:
\bea
sA&=&-W\\
sa&=&-w\nn.
\eea
\end{itemize}
Define left invariant potentials in the usual way:
\bea
	F&=&B-dA\\
	f&=&b-da\nn.
\eea
The set of $SO(2)$ dual actions is then given by (\ref{rotated action}) with:
\bea
	\label{string action}
	S=-\int d^2\s\sqrt{-\mathsf{det}(g_{ij}+F_{ij})}+\int f,
\eea
which differs from that of \cite{hatsuda97} by the inclusion of the $p$-form gauge field and the gauge field rotations. Because of the equivalence of the NS-NS and R-R sectors, all strings in the orbit are viewed as being of the same generalized type. In fact, up to a normalization constant, the actions $S_0$ and $S_\frac{\pi}{4}$ describe the $(1,q)$ and $(p,1)$ elements of the $(p,q)$-strings that are related through the $SL(2,\mathbb{Z})$ duality\footnote{The ``fundamental" string used here has a DBI kinetic term rather than Nambu-Goto. All actions in the $SO(2)$ orbit of the action (\ref{string action}) are D-strings in a generalized sense.} \cite{schwarz97,aganagic97-2,witten95,town97-S-duality,cederwall97}.

Construction of the anomalous term follows along the lines of (\ref{manifest charge algebra}), except that no ``Hodge dual like" fields are required since \textit{both} worldvolume gauge fields are $1$-forms. After constraints are imposed, their conjugate momenta are constants. Define $(P^i,p^i)$ as the doublet of momenta conjugate to $(A_i,a_i)$ respectively. For convenience we define ``hat" and ``check" maps by their action on $(1,n)$-forms $Y,y$:
\bea
	\label{hat check maps for strings}
	\hat Y&=&-\int d\s^1 Y_i P^i\\
	\check y&=&-\int d\s^1 y_i p^i\nn.
\eea
Since the cocycle potentials $(W,w)$ form an $SO(2)$ doublet, and the momenta $(P^i,p^i)$ transform contragradiently, the Noether charges are $SO(2)$ invariant:
\bea
	\bar Q_A&=&\int d\s^1\big (R_A{}^M P_M\big )+\hat W_A +\check w_A.
\eea
Since the Lagrangian is manifestly left invariant, the fully modified charge algebra is then the algebra of Noether charges:
\bea
	\label{string minimal algebra}
	\ [\bar Q_A,\bar Q_B\}&=&-t_{AB}{}^C \bar Q_C+\hat M_{AB}+\check m_{AB},
\eea
which is also $SO(2)$ invariant.

Let us now solve the descent equations to find the anomalous term representatives. Solutions can be obtained by taking linear combinations of all possible terms, and then equating coefficients in the equation. One requires the string Fierz identities:
\bea
	\G_{a(\a\b}\G^a\s_1{}_{\g\d)}&=&0\\
	\G_{a(\a\b}\G^a\s_3{}_{\g\d)}&=&0\nn.
\eea
The first two equations $dB=H$ and $db=h$ are found to be solved by:
\bea
	B&=&\half \Bigg [dx^a-\frac{1}{4}d\bar\t\G^a\t\Bigg ]d\bar\t\G_a\s_1\t\\
	b&=&\half \Bigg [dx^a-\frac{1}{4}d\bar\t\G^a\t\Bigg ]d\bar\t\G_a\s_3\t\nn.
\eea
The next descent equations $sB=-dW$ and $sb=-dw$ then have the solutions:
\bea
	W&=&-\half dx^a\bar\t\G_a\s_1e+\frac{1}{24}d\bar\t\G^a\t\bar\t\G_a\s_1e+\frac{1}{24}\bar\t\G^a ed\bar\t\G_a\s_1\t\\
	w&=&-\half dx^a\bar\t\G_a\s_3e+\frac{1}{24}d\bar\t\G^a\t\bar\t\G_a\s_3e+\frac{1}{24}\bar\t\G^a ed\bar\t\G_a\s_3\t,\nn
\eea
where $e$ refers to the $e^\a$ ghosts. We comment that solutions $b$ and $w$ for type IIB D-branes with higher values of $p$ could be deduced from \cite{kamimura97}. We now obtain the anomalous term representatives $M=sW$ and $m=sw$:
\bea
	\label{string reps for M,m}
	M&=&\half dx^a\bar e\G_a\s_1e+\frac{1}{8}d\Big [\bar e\G^a\t\bar e\G_a\s_1\t\Big ]\\
	m&=&\half dx^a\bar e\G_a\s_3e+\frac{1}{8}d\Big [\bar e\G^a\t\bar e\G_a\s_3\t\Big ].\nn
\eea

Let us now calculate the extended algebras resulting from these representatives. First we need to identify the gauge transformations. These are generated by Lorentz invariant fields in $\Omega^{0,1}$ of dimension two. Define some ``rotated Pauli matrices" as\footnote{The angle $\varphi$ is unrelated to $\phi$ used to rotate the action.}:
\bea
	\s^\varphi_1&=&\cos{(2\varphi)}\s_1-\sin{(2\varphi)}\s_3\\
	\s^\varphi_3&=&\sin{(2\varphi)}\s_1+\cos{(2\varphi)}\s_3\nn.
\eea
By requiring the gauge fields to form a vector doublet $(\Lambda,\lambda)$:
\bea
	\label{IIB lambda gauge fields}
	\Lambda&=&-E x^a\bar e\G_a\s^\varphi_1\t\\
	\lambda&=&-E x^a\bar e\G_a\s^\varphi_3\t\nn,
\eea
the anomalous term remains $SO(2)$ invariant. $E$ and $\varphi$ are free constants which become polar coordinates for the equivalence class of the anomalous term. The gauge transformations generated by (\ref{IIB lambda gauge fields}) are:
\bea
	\label{string delta M and delta m}
	\D M&=&sd\Lambda\\
	&=&-Edx^a\bar e\G_a\s^\varphi_1e-\half Ed\Big [\bar e\G^a\t \bar e\G_a\s^\varphi_1\t\Big ]+Ee^a\bar e\G_a\s^\varphi_1d\t\nn\\
	\D m&=&sd\lambda\nn\\
	&=&-Edx^a\bar e\G_a\s^\varphi_3e-\half Ed\Big [\bar e\G^a\t \bar e\G_a\s^\varphi_3\t\Big ]+Ee^a\bar e\G_a\s^\varphi_3d\t\nn.
\eea
The equivalence classes $[M]$ and $[m]$ are obtained by applying these transformations to the representatives from (\ref{string reps for M,m}). This gives:
\bea
	\label{string anomalous term}
	\ [M]_{\a\b}&=&(1-2E)dx^a(\G_a\s^\varphi_1)_{\a\b}
		+\Bigg [E-\frac{1}{4}\Bigg ]d\Big [(\G^a\t)_{(\a}(\G_a\s^\varphi_1\t)_{\b)}\Big ]\\
	\ [M]_{a\b}&=&-E(\G_a\s^\varphi_1d\t)_\b\nn\\
	\ [m]_{\a\b}&=&(1-2E)dx^a(\G_a\s^\varphi_3)_{\a\b}
		+\Bigg [E-\frac{1}{4}\Bigg ]d\Big [(\G^a\t)_{(\a}(\G_a\s^\varphi_3\t)_{\b)}\Big ]\nn\\
	\ [m]_{a\b}&=&-E(\G_a\s^\varphi_3d\t)_\b\nn.
\eea
One then notes that extended superalgebras are generated from $[M]$ and $[m]$ if the following new generators are defined:
\bea
	\S^a&=&-2dx^a\\
	\S^\a&=&-d\t^\a\nn\\
	\S^\varphi_1{}_{\a\b}&=&-d\Big [(\G^a\t)_{(\a}(\G_a\s^\varphi_1\t)_{\b)}\Big ]\nn\\
	\S^\varphi_3{}_{\a\b}&=&-d\Big [(\G^a\t)_{(\a}(\G_a\s^\varphi_3\t)_{\b)}\Big ]\nn.
\eea
The resulting spectrum of topological charge algebras is then:
\bea
	\label{IIB algebra spectrum}
	\Big \{\bar Q_\a,\bar Q_\b\Big \}&=&-\G^a{}_{\a\b}\bar P_a
		+\Bigg [E-\half\Bigg ]\Big [(\G_a\s^\varphi_1)_{\a\b}\hat \S^a+(\G_a\s^\varphi_3)_{\a\b}\check \S^a\Big  ]\\
		&&-\Bigg [E-\frac{1}{4}\Bigg ]\Big [\hat \S_1^\varphi{}_{\a\b}+\check \S_3^\varphi{}_{\a\b}\Big ]\nn\\
	\Big [\bar Q_\a,\bar P_b\Big ]&=&-E\Big [(\G_b\s^\varphi_1)_{\a\b}\hat\S^\b+(\G_b\s^\varphi_3)_{\a\b}\check\S^\b\Big ]\nn\\
	\Big [\bar Q_\a,\hat\S^b\Big ]&=&-\G^b{}_{\a\b}\hat\S^\b\nn\\
	\Big [\bar Q_\a,\check\S^b\Big ]&=&-\G^b{}_{\a\b}\check\S^\b\nn\\
	\Big [\bar Q_\a,\hat\S^\varphi_1{}_{\b\g}\Big ] &=&\Big [\G^a{}_{\a(\b}(\G_a\s^\varphi_1)_{\g)\d}-\G^a{}_{\d(\b}(\G_a\s^\varphi_1)_{\g)\a}\Big ]\hat\S^\d\nn\\
	\Big [\bar Q_\a,\check\S^\varphi_3{}_{\b\g}\Big ] &=&\Big [\G^a{}_{\a(\b}(\G_a\s^\varphi_3)_{\g)\d}-\G^a{}_{\d(\b}(\G_a\s^\varphi_3)_{\g)\a}\Big ]\check\S^\d\nn.
\eea
The Jacobi identity for the algebra is satisfied due to properties of the cocycle \cite{reimers05}. Indeed, one verifies that the only nontrivial Jacobi identity is given by:
\bea
	\ [\bar Q_\a,\{\bar Q_\b,\bar Q_\g\}]+\mathsf{cycles}&=&\frac{3}{2}\Big [\G^b{}_{(\a\b}(\G_b\s^\varphi_1)_{\g\d)}\hat\S^\d\\
		&&+\G^b{}_{(\a\b}(\G_b\s^\varphi_3)_{\g\d)}\check\S^\d\Big ],\nn
\eea
which vanishes by the Fierz identities.

Only half the fermionic coordinates of the action are physical degrees of freedom due to the presence of $\k$-symmetry. A simple condition one can use to fix $\k$-symmetry is $\t_1=0$ \cite{aganagic97}. In this case $H$ vanishes. It is then simplest to fix the associated potential $B$ and worldvolume gauge field $A$ to be vanishing as well. For simplicity, we will then consider only the ``unbroken" supersymmetries (those preserving $\t_1=0$ without the need for gauge transformations). Under these conditions, the $\S_{\a\b}$ charges vanish, as do all hatted fields and $\check \S^{\a 1}$. The free angular parameter $\phi$ can then be scaled away into $\check \S^a$ and $\check \S^{\a 2}$, and the spectrum reduces to:
\bea
	\label{gauge fixed D string algebra spectrum}
	\Big \{\bar Q_{\a 2},\bar Q_{\b 2}\Big \}&=&-\G^a{}_{\a\b}\bar P_a+\Bigg [E-\half\Bigg ]\G_{a\a\b}\check \S^a\\
	\Big [\bar Q_{\a 2},\bar P_b\Big ]&=&-E\G_{b\a\b}\check\S^{\b 2}\nn\\
	\Big [\bar Q_{\a 2},\check\S^b\Big ]&=&-\G^b{}_{\a\b}\check\S^{\b 2}\nn.
\eea
Due to the gauge condition $\t_1=0$ there is no further equivalence class freedom, so this spectrum is in its most general form. Upon rescaling, it is equivalent to the topological charge algebra derived in \cite{reimers05} of the Green-Schwarz superstring action. This is not surprising since, with the gauge fixing conditions, the $\varphi=0$ action (\ref{string action}) becomes equivalent to the standard Green-Schwarz superstring action \cite{green84}. The only difference is the presence of the $p$-form gauge field in the WZ term, but as in (\ref{manifest charge algebra}) this gauge field has no effect upon the topological charge algebra. The $SO(2)$ rotation $\varphi$ now interpolates between Green-Schwarz and Born-Infeld forms of the action, and this also has no effect upon the charge algebra. The effect that nonlinearly realized supersymmetries of the gauge fixed action have upon the charge algebra is a more complicated problem that we will not address here.

\subsection{$(p,q)$-strings}

To describe $(p,q)$-strings, the action (\ref{string action}) needs modification in order to obtain the required expression for the tension \cite{schwarz97}. We will not explicitly give the required action here (see \cite{town97-S-duality,cederwall97} for a ``duality covariant" formulation). Instead, let us simply note the following properties of the action for a $(J,j)$-string:
\begin{itemize}
\item
The action is manifestly left invariant, and is constructed from the left invariant potentials $(F,f)$.
\item
After constraints are imposed, the momenta $(P^i,p^i)$ conjugate to $(A_i,a_i)$ are:
\bea
	\label{p,q string momenta constraints}
	(P^0,p^0)&=&(0,0)\\
	(P^1,p^1)&=&(J,j)\nn,
\eea
where $(J,j)$ are two integers.
\end{itemize}
This is sufficient information for us to give a topological charge algebra for the $(J,j)$-string. The descent equations once again lead to the representatives (\ref{string reps for M,m}) for the anomalous terms $(M,m)$. The simplest gauge for the resulting algebra is obtained by setting $(E,\varphi)=(\frac{1}{4},0)$ in (\ref{IIB algebra spectrum}). In this case one can remove $\bar \S^\varphi_1{}_{\a\b}$ and $\bar \S^\varphi_3{}_{\a\b}$ from the algebra since they do not appear in the anomalous term. Now impose the constraints (\ref{p,q string momenta constraints}), and factor out the constant momenta from the integrals (\ref{hat check maps for strings}). The algebra is then:
\bea
	\label{p,q string algebra}
	\Big \{\bar Q_\a,\bar Q_\b\Big \}&=&-\G^a{}_{\a\b}\bar P_a
		-\frac{1}{4}\Big [J(\G_a\s_1)_{\a\b}\bar\S^a+j(\G_a\s_3)_{\a\b}\bar\S'^a\Big ]\\
	\Big [\bar Q_\a,\bar P_b\Big ] &=&-\frac{1}{4}\Big [J(\G_b\s_1)_{\a\b}\bar\S^\b+j(\G_b\s_3)_{\a\b}\bar\S'^\b\Big ]\nn\\
	\Big [\bar Q_\a,\bar \S^b\Big ]&=&-\G^b{}_{\a\b}\bar \S^\b\nn\\
	\Big [\bar Q_\a,\bar \S'^b\Big ]&=&-\G^b{}_{\a\b}\bar \S'^\b\nn.
\eea
In the above, we have kept the charges:
\bea
	\bar \S^a&=&\bar \S'^a=2\int d\s^1\del_1 x^a\\
	\bar \S^\a&=&\bar \S'^\a=\int d\s^1\del_1 \t^\a\nn
\eea
distinct, since the general construction allows this. The Jacobi identity:
\bea
	\lsb \bar Q_\a,\{\bar Q_\b,\bar Q_\g\}\rsb+\mathsf{cycles}&=&\frac{3}{2}\Big [J\G^b{}_{(\a\b}(\G_b\s_1)_{\g\d)}\bar\S^\d\\
		&&+j\G^b{}_{(\a\b}(\G_b\s_3)_{\g\d)}\bar\S'^\d\Big ]\nn
\eea
vanishes by the Fierz identities.

These algebras have seen use in the construction of extended superspace actions. The cases $(J,j)=(0,1)$ and $(J,j)=(1,0)$ correspond to algebras used in \cite{sak98,sak00}, while $(J,j)=(1,1)$ corresponds to an algebra used in \cite{sak99}. These algebras can be used to construct left invariant potentials $F$ and $f$ on the associated extended superspaces. This allows extended superspace actions for strings and type IIB D-branes to be constructed. In \cite{reimers05,reimers05-2}, the spectrum of topological charge algebras of standard $p$-brane actions were shown to contain the known algebras that allow the construction of left invariant WZ forms. The appearance of known algebras associated with D-branes in (\ref{p,q string algebra}) generalizes this result. We may observe quite generally that the topological charge algebras generated by a brane cocycle appear to be those which trivialize that cocycle. As a result, these algebras then allow the construction of extended superspace actions.

Note that fermionic winding charges are formally retained and used to close the algebra. The interpretation of fermionic generators as topological charges was considered in \cite{sezgin96}. Such charges are generated, for example, by open strings with different values for fermionic coordinates at the endpoints \cite{hatsuda00}, or by strings bridging a brane-antibrane system \cite{hatsuda01}. Motivation is provided by the fact that fermionic brane charges are necessary in certain backgrounds to ensure quantum consistency with Jacobi identities \cite{peeters03}. In flat backgrounds, the fermionic topological charges have usually been taken to vanish due to the trivial topology associated with fermionic coordinates \cite{dewitt}. In that case, the bosonic charges become ``central" and the entire algebra (\ref{p,q string algebra}) reduces to:
\bea
	\Big \{\bar Q_\a,\bar Q_\b\Big \}&=&-\G^a{}_{\a\b}\bar P_a
		-\frac{1}{4}\Big [J(\G_a\s_1)_{\a\b}\bar\S^a+j(\G_a\s_3)_{\a\b}\bar\S'^a\Big ].
\eea
This type of algebra can be related to partial breaking of rigid supersymmetry \cite{Hughes86-PBRS} via the consideration of particular extended geometries of the brane \cite{sorokin97,townsend97}.

Since $\bar \S^A$ and $\bar \S'^A$ are physically the same charges, a reduced form of the algebra (\ref{p,q string algebra}) can be written where these generators are identified. This is:
\bea
	\label{reduced p,q string algebra}
	\Big \{\bar Q_\a,\bar Q_\b\Big \}&=&-\G^a{}_{\a\b}\bar P_a
		-\frac{1}{4}\Big [J(\G_a\s_1)_{\a\b}+j(\G_a\s_3)_{\a\b}\Big ]\bar\S^a\\
	\Big [\bar Q_\a,\bar P_b\Big ] &=&-\frac{1}{4}\Big [J(\G_b\s_1)_{\a\b}+j(\G_b\s_3)_{\a\b}\Big ]\bar\S^\b\nn\\
	\Big [\bar Q_\a,\bar \S^b\Big ]&=&-\G^b{}_{\a\b}\bar \S^\b\nn.
\eea
Whilst the momenta $(J,j)$ can be viewed as scale factors in (\ref{p,q string algebra}), this is no longer the case in (\ref{reduced p,q string algebra}). The Jacobi identity:
\bea
	\lsb \bar Q_\a,\{\bar Q_\b,\bar Q_\g\}\rsb+\mathsf{cycles}&=&\frac{3}{2}\Big [J\G^b{}_{(\a\b}(\G_b\s_1)_{\g\d)}\\
		&&+j\G^b{}_{(\a\b}(\G_b\s_3)_{\g\d)}\Big ]\bar\S^\d\nn
\eea
again vanishes.

\section{Application to the D-membrane}
\label{sec:Application to the D-membrane}

Let us solve the descent equations for the D2-brane in order to find representatives for the two anomalous terms of the Noether charge algebra. The Fierz identities for the membrane are required:
\bea
	\G^a{}_{(\a\b}(\G_{11}\G_a)_{\g\d)}&=&0\\
	\G_{11(\a\b}(\G_{11}\G_a)_{\g\d)}-\G^b{}_{(\a\b}\G_{ab\g\d)}&=&0\nn.
\eea
We begin with the NS-NS sequence. The solution for $B$ is found to be:
\bea
	B&=&\half \Bigg [dx^a-\frac{1}{4}d\bar\t\G^a\t\Bigg ]d\bar\t\G_{11}\G_a\t.
\eea
The equation $sB=-dW$ is then solved by\footnote{An analogous solution (without ghost fields) appears in \cite{hammer97}.}:
\bea
	W&=&-\half dx^a\bar\t\G_{11}\G_ae+\frac{1}{24}d\bar\t\G^a\t\bar\t\G_{11}\G_a e+\frac{1}{24}\bar\t\G^a ed\bar\t\G_{11}\G_a \t.
\eea
This yields the representative $M=sW$ for the NS-NS anomalous term:
\bea
	\label{membrane M}
	M&=&\half dx^a\bar e\G_{11}\G_a e+\frac{1}{8}d\Big [\bar e\G^a\t\bar e\G_{11}\G_a\t\Big ].
\eea
We now turn to the WZ cocycle. One might deduce representatives for $b$ and $w$ for type IIA D-branes from \cite{hatsuda98}, however to illustrate the procedure we will find these quantities for the D2-brane. The equation for $\breve b^{(1)}$ that follows from (\ref{R-R potential defn}) is:
\bea
	d\breve b^{(1)}&=&R^{(2)},
\eea
which is easily solved by:
\bea
	\breve b^{(1)}&=&\half d\bar\t\G_{11}\t.
\eea
The equation for $\breve b^{(3)}$ is then:
\bea
	d\breve b^{(3)}&=&R^{(4)}-\breve b^{(1)}H.
\eea
This is solved by:
\bea
	\breve b^{(3)}&=&\frac{1}{4}dx^a dx^b d\bar\t\G_{ab}\t\\
		&&+dx^a\Bigg [-\frac{1}{8}d\bar\t\G^b\t d\bar\t\G_{ab}\t+\frac{1}{8}d\bar\t \G_{11}\t d\bar\t \G_{11}\G_a \t\Bigg  ]\nn\\
		&&+d\bar\t\G^a\t \Bigg [\frac{1}{48}d\bar\t \G^b\t d\bar\t\G_{ab}\t-\frac{1}{24}d\bar\t\G_{11}\t d\bar\t\G_{11}\G_a\t\Bigg ]\nn.
\eea
From (\ref{sb descent equation}) we determine the equation for $\breve w^{(0)}$:
\bea
	d\breve w^{(0)}&=&-s\breve b^{(1)},
\eea
which is easily solved by:
\bea
	\breve w^{(0)}&=&-\half \bar\t\G_{11}e.
\eea
The equation for $\breve w^{(2)}$ is then:
\bea
	d\breve w^{(2)}&=&-s\breve b^{(3)}-\breve w^{(0)}H.
\eea
This is solved by:
\bea
	\breve w^{(2)}&=&-\frac{1}{4}dx^adx^b\bar\t\G_{ab}e\\
		&&+\frac{1}{24}dx^a\Big [\bar\t\G^b ed\bar\t\G_{ab}\t+d\bar\t\G^b\t \bar\t\G_{ab}e+5\bar\t\G_{11}ed\bar\t\G_{11}\G_a \t\nn\\
		&&-d\bar\t\G_{11}\t \bar\t\G_{11}\G_a e\Big ]\nn\\
		&&+\frac{1}{240}\Big [-d\bar\t\G^a\t d\bar\t\G^b\t \bar\t\G_{ab}e+\bar\t\G^a ed\bar\t\G^b\t d\bar\t\G_{ab}\t\nn\\
	&&+2d\bar\t\G^a\t d\bar\t\G_{11}\t \bar\t\G_{11}\G_a e-14d\bar\t\G^a\t \bar\t\G_{11}ed\bar\t\G_{11}\G_a \t\nn\\
		&&-\bar\t\G^ae d\bar\t\G_{11}\t d\bar\t\G_{11}\G_a \t\Big ]\nn.
\eea
We then finally obtain the forms:
\bea
	\breve m^{(0)}&=&\half \bar e\G_{11}e\\
	\breve m^{(2)}&=&-\frac{1}{4}dx^adx^b\bar e\G_{ab}e\nn\\
		&&+dx^a\Bigg [\frac{1}{24}\bar\t\G^be d\bar\t\G_{ab}e-\frac{1}{24}\bar e\G^be d\bar\t\G_{ab}\t-\frac{1}{24}d\bar\t\G^b\t e\G_{ab}e\nn\\
		&&-\frac{7}{24}d\bar\t\G^be \bar\t\G_{ab}e+\frac{5}{24}\bar\t\G_{11}e d\bar\t\G_{11}\G_ae-\frac{5}{24}\bar e\G_{11}e d\bar\t\G_{11}\G_a\t\nn\\
		&&+\frac{1}{24}d\bar\t\G_{11}\t \bar e\G_{11}\G_ae+\frac{1}{24}d\bar\t\G_{11}e \bar\t\G_{11}\G_ae\Bigg ]\nn\\
		&&+\frac{1}{240}d\bar\t\G^a\t d\bar\t\G^b\t \bar e\G_{ab}e-\frac{1}{80}d\bar\t\G^a\t d\bar\t\G^be \bar\t\G_{ab}e\nn\\
		&&+\frac{1}{240}\bar\t\G^ae d\bar\t\G^b\t \bar d\t\G_{ab}e-\frac{1}{60}\bar\t\G^ae d\bar\t\G^be \bar d\t\G_{ab}\t\nn\\
		&&-\frac{1}{240}\bar e\G^ae d\bar\t\G^b\t \bar d\t\G_{ab}\t-\frac{1}{120}d\bar\t\G^a\t d\bar\t\G_{11}\t \bar e\G_{11}\G_ae\nn\\
		&&-\frac{1}{120}d\bar\t\G^a\t d\bar\t\G_{11}e \bar \t\G_{11}\G_ae+\frac{1}{80}d\bar\t\G^ae d\bar\t\G_{11}\t \bar \t\G_{11}\G_ae\nn\\
		&&-\frac{7}{120}d\bar\t\G^a\t \bar\t\G_{11}e d\bar \t\G_{11}\G_ae+\frac{7}{120}d\bar\t\G^a\t \bar e\G_{11}e d\bar \t\G_{11}\G_a\t\nn\\
		&&-\frac{11}{240}d\bar\t\G^ae \bar\t\G_{11}e d\bar \t\G_{11}\G_a\t-\frac{1}{240}\bar\t\G^ae d\bar\t\G_{11}\t d\bar \t\G_{11}\G_ae\nn\\
		&&-\frac{1}{240}\bar\t\G^ae d\bar\t\G_{11}e d\bar \t\G_{11}\G_a\t+\frac{1}{240}\bar e\G^ae d\bar\t\G_{11}\t d\bar \t\G_{11}\G_a\t\nn.
\eea
The corresponding representative of the WZ anomalous term is given by $\bar m$, where:
\bea
	\label{WZ anomalous term}
	m&=&\breve m^{(0)}F+{\breve m^{(2)}}.
\eea
The first term contains a topological integral of the field strength of the BI gauge field, while the first term of $\breve m^{(2)}$ is a familiar bosonic term:
\bea
	dx^adx^b\bar e\G_{ab}e
\eea
that also exists in the case of ordinary $p$-branes \cite{azc89}. These two terms, plus the first term of (\ref{membrane M}), generate the three central extensions of the standard supertranslation algebra that are associated with bosonic topology \cite{hammer97}. The remaining terms are the ones associated with fermionic topology which generalize the solutions of \cite{hammer97,hatsuda98}. Since the number of fermionic terms is quite large, we will not explicitly calculate the spectrum of algebras associated with the WZ anomalous term (\ref{WZ anomalous term}) in this work.

Let us now calculate the extended algebras resulting from the NS-NS anomalous term. Two Lorentz invariant $\Lambda$ gauge fields with the correct dimensionality are:
\bea
	\label{IIA lambda gauge fields}
	\Lambda_1&=&-x^a\bar e\G_a\t\\
	\Lambda_2&=&-x^a\bar e\G_{11}\G_a\t\nn.
\eea
A third possibility:
\bea
	\Lambda_3&=&-2e^ax^b\eta_{ab}
\eea
is equivalent to $\Lambda_1$ since $sd\Lambda_3=sd\Lambda_1$. Some other possibilities that we will not use here are given in appendix \ref{sec:Additional gauge fields}. The gauge transformations generated are:
\bea
	\D_1M&=&sd\Lambda_1\\
	&=&-dx^a\bar e\G_ae+e^a\bar e\G_ad\t\nn\\
	\D_2M&=&sd\Lambda_2\nn\\
	&=&-dx^a\bar e\G_{11}\G_ae-\half d\Big [\bar e\G^a\t \bar e\G_{11}\G_a\t\Big ]+e^a\bar e\G_{11}\G_ad\t\nn.
\eea
The full class [M] for the anomalous term is then obtained by applying these transformations to the representative (\ref{membrane M}):
\bea
	[M]&=&M+E_1\D_1 M+E_2\D_2 M,
\eea
where $E_1$ and $E_2$ are free constants which parameterize the class. This gives:
\bea
	\ [M]_{\a\b}&=&(1-2E_2)dx^a(\G_{11}\G_a)_{\a\b}-2E_1dx^a\G_{a\a\b}\\
		&&+\Bigg [E_2-\frac{1}{4}\Bigg ]d\Big [(\G^a\t)_{(\a}(\G_{11}\G_a\t)_{\b)}\Big ]\nn\\
	\ [M]_{a\b}&=&-E_1(\G_ad\t)_\b-E_2(\G_{11}\G_ad\t)_\b\nn.
\eea
One then notes that $[M]$ generates a spectrum of extended superalgebras if three new generators are defined\footnote{A term analogous to $\S_{\a\b}$ was obtained in \cite{hatsuda98}. However, due to the trivial fermionic topology used there, a vanishing charge was obtained. One also does not obtain $[Q,P]$ or $[P,P]$ anomalous terms under this assumption since such charges are fermionic on dimensional grounds (see \cite{reimers05-2}).}:
\bea
	\S^a&=&-2dx^a\\
	\S^\a&=&-d\t^\a\nn\\
	\S_{\a\b}&=&-d\Big [(\G^a\t)_{(\a}(\G_{11}\G_a\t)_{\b)}\Big ]\nn.
\eea
The resulting topological charge algebra is then:
\bea
	\label{IIA algebra spectrum}
	\Big \{\widetilde{\bar Q}_\a,\widetilde{\bar Q}_\b\Big \}&=&-\G^a{}_{\a\b}\widetilde{\bar P}_a+\lsb\Bigg [E_2-\half\Bigg ](\G_{11}\G_a)_{\a\b}+E_1\G_{a\a\b}\rsb\hat \S^a\\
		&&-\Bigg [E_2-\frac{1}{4}\Bigg ]\hat \S_{\a\b}\nn\\
	\Big [\widetilde{\bar Q}_\a,\widetilde{\bar P}_b\Big ]&=&-\Big [E_1\G_{b\a\b}+E_2(\G_{11}\G_b)_{\a\b}\Big ]\hat \S^\b\nn\\
	\Big [\widetilde{\bar Q}_\a,\hat\S^b\Big ]&=&-\G^b{}_{\a\b}\hat \S^\b\nn\\
	\Big [\widetilde{\bar Q}_\a,\hat\S_{\b\g}\Big ]&=&\Big [\G^a{}_{\a(\b}(\G_{11}\G_a)_{\g)\d}-\G^a{}_{\d(\b}(\G_{11}\G_a)_{\g)\a}\Big ]\hat \S^\d\nn.
\eea
The Jacobi identity for the algebra is again satisfied due to properties of the cocycle. One verifies that the only nontrivial possibility is:
\bea
	\Big [\widetilde{\bar Q}_\a,\Big \{\widetilde{\bar Q}_\b,\widetilde{\bar Q}_\g\Big \}\Big ]+\mathsf{cycles}&=&\frac{3}{2}\G^b{}_{(\a\b}(\G_{11}\G_b)_{\g\d)}\hat\S^\d,
\eea
which vanishes by the standard Fierz identity. The algebra (\ref{IIA algebra spectrum}) is not restricted to the membrane. Apart from the worldvolume embedding, the DBI term of the D$p$-brane action is the same for all values of $p$. The NS cocycle thus generates the same algebra for all standard, type IIA D-brane actions with $p\geq 2$. Similarly, the subalgebra of (\ref{IIB algebra spectrum}) which contains only the NS charges is the same for all type IIB D-branes.

An algebra within the spectrum (\ref{IIA algebra spectrum}) has also been used in the context of trivializing cocycles. In the special case $E_2=\frac{1}{4}$, $\S_{\a\b}$ is not present in the anomalous term and can be excluded from the algebra. The gauge $E_1=0$ then yields an algebra which corresponds to one used in the construction of extended, type IIA superspace actions for strings, D-branes and string-brane systems \cite{sak98,chrys99,hatsuda01}\footnote{A redefinition $\S^\a\rightarrow \G_{11}\S^\a$ is required to establish the correspondence with \cite{sak98,hatsuda01}.}. We note that in both the type IIA and IIB cases, the free constants in the spectra do not correspond to rescalings of the previously known algebras. The Noether charge algebras of standard superspace D-brane actions thus generate new candidates for the algebras underlying extended superspace action formulations.

\section{Comments}
\label{sec:Comments}

Recently we have been investigating topological charge algebras associated with brane cocycles. We find that these algebras are such that they allow the trivialization of the cocycle from which they derive. As a result, in the case of $p$-branes, the algebras allow the construction of left invariant WZ forms. For D-branes, they additionally allow the construction of extended superspace actions without worldvolume gauge fields. Such actions have already been constructed using previously known algebras \cite{sak98,sak99,chrys99,sak00,hatsuda01}. We would like to determine whether all algebras in the spectra derived in this paper can be used to construct extended superspace actions. Work on this issue is in preparation.

For simplicity, we have here considered actions without the background scalars. If these scalars are included, the action is invariant when they take their vacuum values. Representatives for the required anomalous terms then result from solving the same descent equations. The process thus depends only upon the field strengths (i.e. nontrivial cocycles) involved, and the background scalars do not contribute directly to the topological charge algebra. However, they may contribute indirectly through the consideration of dualities (for example, as a restriction on the gauge fields, as in section \ref{sec:D-strings}). We note here that an algebra parameterized by the background scalars was considered in \cite{sak99}. This type of algebra might be expected to arise as a topological charge algebra if the scalars (belonging to the coset $SL(2,\mathbb{R})/SO(2)$) are identified with coordinates of the duality group.

\subsection{Acknowledgments}
I would like to thank I. N. McArthur for helpful suggestions and critical reading of the manuscript.

\appendix
\section{Appendix}

\subsection{Additional gauge fields}
\label{sec:Additional gauge fields}

The gauge transformations (\ref{IIB lambda gauge fields}) and (\ref{IIA lambda gauge fields}) are not the only ones consistent with dimensionality and Lorentz invariance. For example, in the IIA case one can also consider the gauge fields:
\bea
	\Lambda^{(b)}&=&\bar e\G^{a_1\ldots a_b}\t \bar\t\G_{11}\G_{a_1\ldots a_b}\t\\
	\Lambda'^{(b)}&=&\bar e\G_{11}\G^{a_1\ldots a_b}\t \bar\t\G_{a_1\ldots a_b}\t\nn,
\eea
where in $\Lambda^{(b)}$, $b$ is such that $\G_{11}\G_{a_1\ldots a_b}$ is antisymmetric, while in $\Lambda'^{(b)}$, $b$ is such that $\G_{a_1\ldots a_b}$ is antisymmetric. The minimal Green-Schwarz superstring action appears to be special in that this type of gauge transformation does not contribute to the topological charge algebra \cite{reimers05}. In the present type IIA example extra terms are contributed to (\ref{IIA algebra spectrum}), however there are no extra generators required. Define $\D^b M=sd\Lambda^{(b)}$ and $\D'^b M=sd\Lambda'^{(b)}$. For $E_2\neq \frac{1}{4}$ one can then set:
\bea
	\S'_{\a\b}&=&\S_{\a\b}-\lsb\frac{1}{E_2-\frac{1}{4}}\rsb (E_b\D^b M_{\a\b}+E'_b\D'^b M_{\a\b}).
\eea
The only alteration to the algebra then occurs as additional terms on the RHS of $[Q_\a,\S_{\b\g}]$. These additional terms do not appear to contribute to calculations involving trivialization of the cocycle (a point we will not illustrate here). Since this appears to be the main application of the algebras, we chose not to make use of such gauge transformations.

\bibliographystyle{hieeetr}
\bibliography{double_complex}
\end{document}